\documentclass[11pt,twoside]{article}
\pdfoutput=1
\usepackage{asp2014}

\aspSuppressVolSlug
\resetcounters

\begin{document}

\title{An Automated Pipeline for the VST Data Log Analysis}

\author{S.~Savarese, P.~Schipani, G.~Capasso, M.~Colapietro, S.~D'Orsi, L.~Marty, and F.~Perrotta}
\affil{INAF Osservatorio Astronomico di Capodimonte, Salita Moiariello 16, Naples, Italy
\\\email{salvatore.savarese@inaf.it}}

\paperauthor{Salvatore~Savarese}{salvatore.savarese@inaf.it}{https://orcid.org/0000-0002-4778-6050}{INAF Osservatorio Astronomico di Capodimonte}{}{Naples}{}{80131}{Italy}
\paperauthor{Pietro~Schipani}{pietro.schipani@inaf.it}{}{INAF Osservatorio Astronomico di Capodimonte}{}{Naples}{}{80131}{Italy}
\paperauthor{Giulio~Capasso}{giulio.capasso@inaf.it}{}{INAF Osservatorio Astronomico di Capodimonte}{}{Naples}{}{80131}{Italy}
\paperauthor{Mirko~Colapietro}{mirko.colapietro@inaf.it}{}{INAF Osservatorio Astronomico di Capodimonte}{}{Naples}{}{80131}{Italy}
\paperauthor{Sergio~D'Orsi}{sergio.dorsi@inaf.it}{}{INAF Osservatorio Astronomico di Capodimonte}{}{Naples}{}{80131}{Italy}
\paperauthor{Laurent~Marty}{laurent.marty@inaf.it}{}{INAF Osservatorio Astronomico di Capodimonte}{}{Naples}{}{80131}{Italy}
\paperauthor{Francesco~Perrotta}{francesco.perrotta@inaf.it}{}{INAF Osservatorio Astronomico di Capodimonte}{}{Naples}{}{80131}{Italy}
  
\begin{abstract}

The VST Telescope Control Software logs continuously detailed information \linebreak about the telescope and instrument operations. Commands, telemetries, errors, weather conditions and anything may be relevant for the instrument maintenance and the identification of problem sources is regularly saved. All information are recorded in textual form.
These log files are often examined individually by the observatory personnel for specific issues and for tackling problems raised during the night. Thus, only a minimal part of the information is normally used for daily maintenance.
Nevertheless, the analysis of the archived information collected over a long time span can be exploited to reveal useful trends and statistics about the telescope, which would otherwise be overlooked. Given the large size of the archive, a manual inspection and handling of the logs is cumbersome. An automated tool with an adequate user interface has been developed to scrape specific entries within the log files, process the data and display it in a comprehensible way. This pipeline has been used to scan the information collected over 5 years of telescope activity.
  
\end{abstract}

\section{Introduction}

The 2.6-m VST is one of the telescopes of the ESO Paranal Observatory in the Chilean Atacama desert. The VST control software consists of many parallel processes running on different machines, all  writing information to log files. The information may be related to normal operations (e.g. the execution of commands is typically logged whenever they are executed, with all the relevant parameters), unforeseen events (e.g. errors), debug activity (useful during software development and telescope commissioning). The logs can be periodic (e.g. data coming from the weather station) or asynchronous, like commands triggered by the users' activity.
All the information about the telescope and instrument operations is recorded in these files, which are automatically mirrored to the ESO Science Archive Facility that hosts the science data of all the instruments of the observatory. This makes the operational data remotely available through the same well established procedures used by the ESO community astronomers to access the scientific data. 

A Matlab pipeline has been developed for the remote analysis of operational data. The main purposes of this work are a remote monitoring and the ability to perform statistical studies \citep{Schipani2020}, for which a daily rate is not needed. In fact, a prompt analysis and immediate tracking of issues makes sense only for the observatory staff members, who work on site. Instead, for a remote analysis, having all the files available on monthly basis is enough. Thus the log files are retrieved with a period of one month and analyzed in batches.

\articlefigure[width=.9\textwidth]{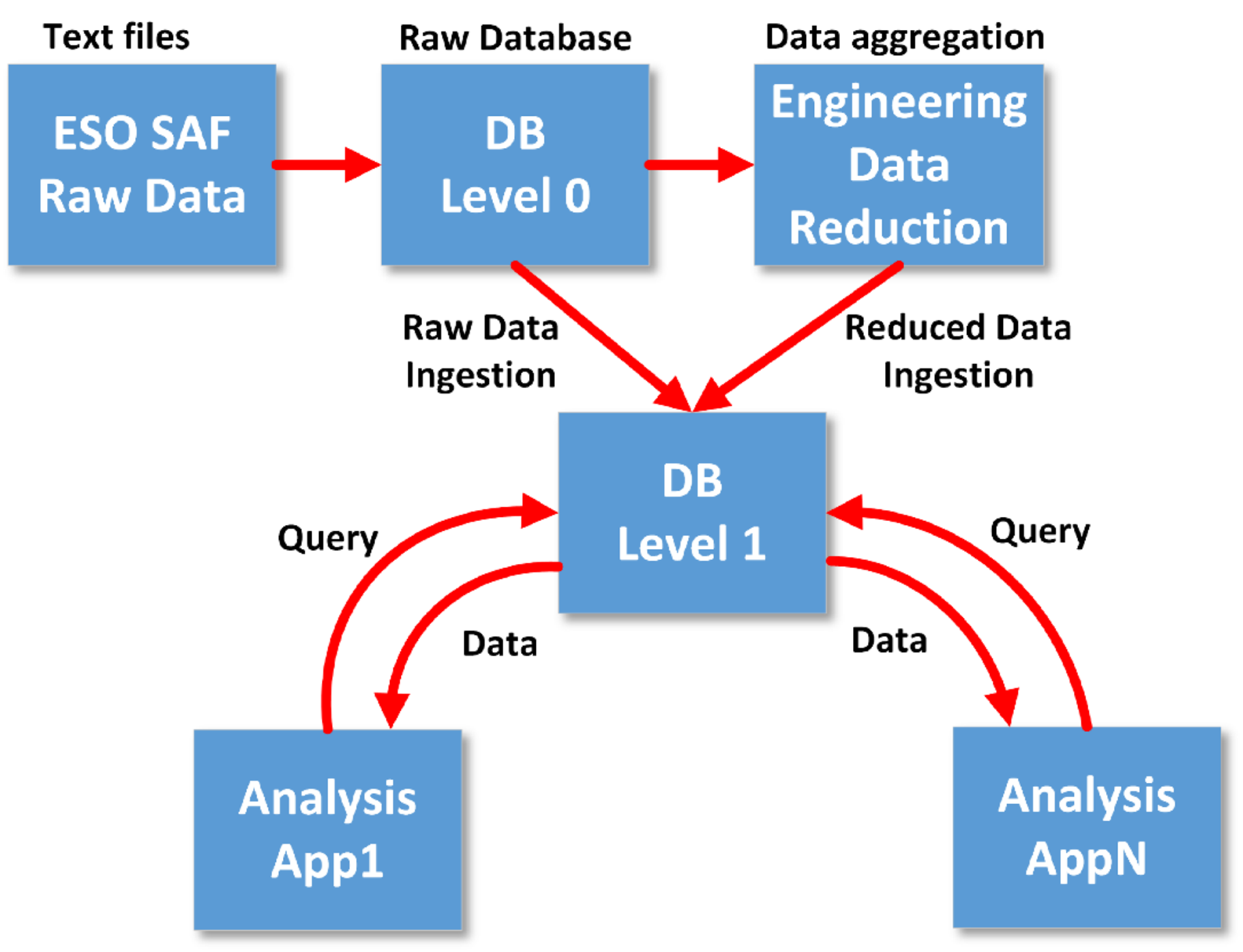}{fig1}{Block diagram of the pipeline. The raw data are extracted and then reduced by the pipeline, creating an enriched database which is queried by specific applications for the analysis.}

\section{Pipeline}
The data obtained from the ESO archive are processed in three steps, summarized in the next sections and represented by the data flow in Figure \ref{fig1}.

\subsection{File Scanning}
The first step of the elaboration is parsing the text files to extract and sort the lines corresponding to events and commands of interest. Each line consists of a fixed length timestamp, followed by a text field describing the event, sometimes terminated by numerical data. The latter is optional, and can range from a single number like the reading of a sensor, to multiple values (e.g. five for the position of the secondary mirror, three spatial and two rotational coordinates). Instead, some events are completely described by their timestamp, such as the start of a new exposure or, more in general, system commands.
The search is customized with a configuration file wherein the user defines which data must be extracted from the log. Each line of the configuration file corresponds to a specific event or command and consists of the following fields:

\begin{itemize}
	\item Search Pattern, i.e. one or more strings that uniquely identify the command;
	\item Name to assign to the corresponding variable in MATLAB;
	\item Range of values;
	\item The format of the data (numeric or text).
\end{itemize}

The configuration file can be easily customized for searching new patterns, thus extending the contents of the output database. 
For each defined  entry, a structure array is created to contain its data and timestamps in the MATLAB workspace.
The parsing is the only time consuming step. If new patterns are added, the rescan of the the whole archive is needed in order to maintain a coherent set of data. Luckily, the log files feature a simple and consistent formatting, not requiring overly complex search patterns like regular expressions, which would slow down the process. 

The scan is performed just to build the database where the data are stored. Once the database is extracted, the data size to be handled is significantly reduced from several GB for the whole archive to few hundreds MB.

\subsection{Data Reduction}
After the log files have been parsed, a database of raw data (``Level 0" DB) is available, plenty of information but still not ready for any possible analysis. A data reduction phase is needed to complement the database with ``reduced" data, assuming for this engineering data processing the same terminology adopted for science data pipelines. The reduced data will then be ingested to the database, complementing the raw data coming from the scan. 

To make an example of data reduction within this context,  a case study for the analysis of the Active Optics system \citep{Schipani2016} is considered. A typical Active Optics cycle is composed of many separate events and data, which must be correctly aggregated to give a complete picture of the whole process, because no single log contains all the data which are needed for the analysis. The data reduction must guarantee that the aggregation of the events is correct. For instance, to study the behaviour of the open-loop corrections, the block of events between the open-loop correction and the first exposure of an Observation Block is all relevant. They must be isolated and aggregated, building a set of meaningful data referred to any individual open-loop correction. These data are aggregated from the raw tables, mainly working on the timestamps and the knowledge of the underlying logic of the telescope software. 

This logic is represented in Figure \ref{fig2}, showing the timeline for a typical Observation Block. The commands executed before the exposure are the following: first a {\it preset} command is issued to slew to a new target; when the telescope reaches the target the {\it tracking} phase starts, which usually, but not always, triggers a {\it onecal} command to execute a calibrated, open-loop, Active Optics  correction; then few closed-loop Active Optics corrections ({\it IA} - Image Analysis) are executed to remove any residual aberration; finally, the sequence of exposures (yellow blocks) begins, starting the AutoGuider and stopping it before applying the offsets for dithering. Further closed-loop corrections of the active optics system are applied between the exposures. 

Thus, the open-loop command and the first nearby exposure define an interval containing the next closed-loop corrections, which give information on the quality of the open-loop look-up table. The data reduction software aggregates the data, building additional information which are then ingested to the database as ``Level 1" products.

\articlefigure{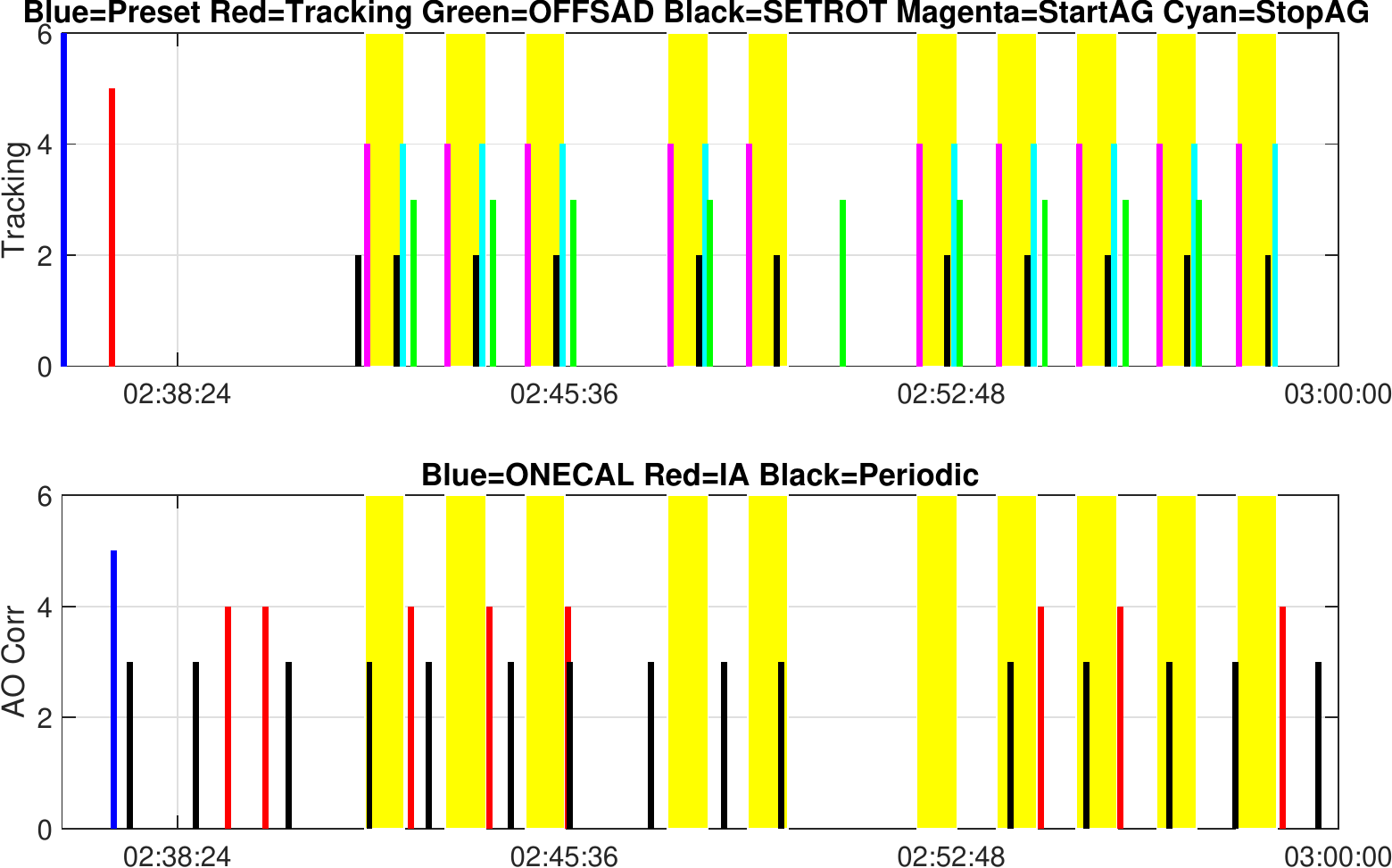}{fig2}{Command sequence for a typical Observation Block, with exposure times (open shutter) in yellow. Top: events related to the tracking system. Bottom: events related to the active optics system.}

\subsection{Analysis of data}
Once the ``Level 1" database is populated with reduced data products, many more aspects of the system can be studied. Of course, the analysis phase is tightly correlated with the data reduction: the analysis is possible only for data which have been prepared and stored into the database. Nevertheless, the process is modular and incremental. New reduced data aggregations can be added at any time, triggering the development of different apps (see Figure \ref{fig1}) for making specific new analysis.

\section{Conclusions}
A pipeline for the investigation of the VST telescope logs has been described. It is available for statistical studies or for analysis of sporadic night issues. This tool may be helpful to identify hidden issues and further improve the telescope operations.

\bibliography{P11-208}

\end{document}